\documentclass[12pt,preprint]{aastex}
\usepackage{amsmath}
\usepackage{lscape}
\usepackage{comment}
\begin{document}

\title{Intrinsically Polarized Stars and Implication for Star Formation\\ in the Central Parsec of Our Galaxy}
\author{Tatsuhito Yoshikawa\altaffilmark{1}, Shogo Nishiyama\altaffilmark{2}, Motohide Tamura\altaffilmark{2}, 
Miki Ishii\altaffilmark{3} \\ and Tetsuya Nagata\altaffilmark{1}}
\altaffiltext{1}{Department of Astronomy, Graduate School of Science, Kyoto University, Kyoto 606-8502}
\altaffiltext{2}{National Astronomical Observatory of Japan, Mitaka, Tokyo 181-8588}
\altaffiltext{3}{Subaru telescope, National Astronomical Observatory of Japan, 650 North A'ohoku Place, Hilo, HI 96720, USA}
\email{yosikawa@kusastro.kyoto-u.ac.jp}
%\date{}
%\maketitle

%%%%%%%%%%%%%%%%%%abstract%%%%%%%%%%%%%%%%%%%%
\begin{abstract} 
We have carried out adaptive-optics assisted observations at the Subaru telescope, 
and have found 11 intrinsically polarized sources in the central parsec of our Galaxy. 
They are selected from 318 point sources with $K_S<15.5$, 
and their interstellar polarizations are corrected using a Stokes $Q/I-U/I$ diagram. 
Considering brightness, near-infrared color excess, and the amount of intrinsic polarization,  
two of them are good young stellar object (YSO) candidates with an age of $\sim10^5$\,yr. 
If they are genuine YSOs, their existence provides strong constraints on star formation mechanisms in this region. 
In the remaining sources, two are known as bow-shock sources in the Northern arm. 
One other is also located in the Northern arm and shows very similar properties, 
and thus likely to be a so far unknown bow-shock source.
The origin of the intrinsic polarization of the other sources is as yet uncertain.
\end{abstract}
%%%%%%%%%%%%%%%%%%%%abstract%%%%%%%%%%%%%%%%%%%%
\keywords{stars: formation --- stars: pre-main sequence --- Galaxy: center --- polarization}

%%%%%%%%%%%%%%%%%%%%introduction%%%%%%%%%%%%%%%%%%%%
\section{Introduction}
In the central parsec of our Galaxy, more than 100 young massive stars exist. 
These include helium-rich blue supergiants, Wolf-Rayet stars with ZAMS masses up to $\sim$100M$_{{\sun}}$, 
and OB main sequcence stars \citep{kra95,mou05,pau06,bar09,bar10}.
However, the strong tidal force from the supermassive black hole (SMBH) Sgr A* makes it very difficult for stars
to be formed by gravitational collapse of a molecular cloud \citep{mor93}.
Investigating how the young massive stars were formed is 
important to understand the star formation process under the strong gravitational field from the SMBH. 

Various hypotheses have been constructed to understand the presence of young massive stars near the SMBH 
\citep[e.g.,][]{ale05,gen10} with two of them currently favoured:
the ``in situ, accretion disk'' scenario \citep[e.g.,][]{lev03,gen03} and
the ``inspiraling star cluster'' scenario \citep[e.g.,][]{ger01,kim03}.
The former scenario is star formation in situ ($\lesssim$1\,pc from the SMBH) in a massive self-gravitating disk, 
formed by the infall of a large molecular cloud.
Several numerical simulations have shown that this mechanism works well \citep{bon08,hob09,ali09}.
In the latter scenario, ``normal'' star formation occurs in a very massive star cluster tens of parsecs from the SMBH, 
and then the stars spiraled into the central-parsec region as a result of angular momentum loss through dynamical friction.

One method to place strong constraints on the latter scenario is to find younger stars in the central parsec, 
because such stars would have to be able to infall into the central parsec within their current lifetime.
Recent observations using adaptive optics (AO) suggest
that the young massive stars now observed were formed in a starburst 6$\pm$2\,Myr ago \citep{pau06,bar09,lu13}.
When we assume the initial distance of a star cluster to be 30\,pc from the SMBH,
the required cluster mass is $\sim10^6$ M$_{{\sun}}$ to infall to the central parsec
within 6\,Myr \citep[see Equation 8 in][]{ger01}.
Although young clusters (Quintuplet and Arches) exist at the distances of $\sim$30\,pc \citep[e.g.,][]{oku90,nag95,fig04}, 
a cluster mass of $10^6$\,M$_{{\sun}}$ is two-orders greater than the masses of these clusters,
and as massive as the largest globular clusters.
If we find stars younger than 1\,Myr, such as young stellar objects (YSOs) with circumstellar disks, 
we can place more stringent constraints on the star formation scenario in the central parsec region.

Various papers have claimed the detection of YSO candidates in the Galactic center environment.
\cite{cle01} and \cite{eck04} construct color-magnitude diagrams with the $K$ and $L$ bands to identify individual sources.
\cite{eck04} indicate the possibility that IRS\,13N, which is a complex of extremely red sources, 
consists of Herbig Ae/Be objects with ages of about 0.1 to 1\,Myr. 
\cite{muz08} and \cite{eck13} carry out astrometric studies of IRS\,13N, 
and their results favor the scenario which IRS\,13N is a complex of YSOs.
\cite{yus13} find 11 SiO (5-4) clumps of molecular gas within 0.6\,pc of Sgr A* using ALMA (Atacama Large Millimeter/submillimeter Array) data
and they interpret them as highly embedded protostellar outflows, 
signifying an early stage of massive star formation in the last $10^4-10^5$\,yr.

In this study, we use another method to detect YSOs in the Galactic center.
To find YSOs with a wide range of masses, linear polarimetric observations provide effective information
because many YSOs are intrinsically polarized due to the scattering of the stellar light by dust grains in their circumstellar disk.
This is confirmed in model calculations \citep{whi92} and observations \citep{tam89,tam05,yud00,per09}.
The age of Herbig Ae/Be stars, which are intermediate mass pre-main sequence stars with circumstellar disks,
are typically younger than $10^6$\,yr \citep{alo09}
and bright enough for a detection at the distance of the Galactic center \citep[e.g.,][]{eir02}.
Therefore the purpose of our study is to find Herbig Ae/Be stars in the central 1-parsec region through polarimetric observations.

Toward the central parsec, polarimetric observations are sparse; 
the newest one covering the entire parsec 
was carried out by \cite{ott99}, whose limiting magnitude is 11 in the $K$ band, 
although a much deeper observation was recently made in two strips \citep{buc11}.
A few intrinsically polarized stars such as IRS\,21 were reported, 
and they are most likely to be bow shock sources \citep[e.g.,][and see Section 4.2]{tan02,tan05}.
We carried out deep polarimetric observations of the central parsec region to search for YSOs using their intrinsic polarization.
%%%%%%%%%%%%%%%%%%%%introduction%%%%%%%%%%%%%%%%%%%%

%%%%%%%%%%%%%%%%%%%%observation%%%%%%%%%%%%%%%%%%%%
\section{Observations of polarization in the central parsec}
We carried out $K_S$-band (2.15\,$\mu$m) polarimetric observations
using CIAO (Coronagraph Imager with Adaptive Optics) and its polarimeter \citep{tam00,tam03} with AO36 \citep{tak04}
on the Subaru telescope\footnote{Based on data collected at Subaru Telescope, 
which is operated by the National Astronomical Observatory of Japan.} \citep{iye04} on 2007 May 26 - 28.
CIAO provides an image of a 22$''$.2 $\times$ 22$''$.2 area of the sky with a scale of 21.7\,mas pix$^{-1}$.
With the $R = 13.7$\,mag natural guide star USNO 0600-28577051 located $\approx$20$''$ from Sgr A*,
and stable atmospheric conditions during the observations,
AO36 provided stable correction with seeing values between 0$''$.15 - 0$''$.25 in the $K_S$ band through three nights. 

We observed a field of 22$''$.2 $\times$ 22$''$.2 of the Galactic center centering Sgr A* for three nights.
This corresponds to a square of $\sim$ 0.9\,pc
at a distance to the Galactic center of 8\,kpc \citep[e.g., ][]{gil09}.
To obtain polarization, we used a rotating half-wave plate with a fixed wire grid analyzer.
We made 20\,sec exposures at four waveplate angles in the sequence of
0$^\circ$, 45$^\circ$, 22.5$^\circ$, and 67.5$^\circ$ (1\,set)
and we carried out 189\,sets of observations during three nights.
For checking reproducibility (see 3.1),
we combined our data by each night and obtained three-one-night data sets.
The integration time is 800\,sec for the first night, 1140\,sec for the second night, and 1500\,sec for the third night;
we removed 17 low-quality data sets. We processed the observed data using IRAF/DAOPHOT in a normal manner
- dark subtraction, flat field correction, sky subtraction, and dead pixel correction.
%%%%%%%%%%%%%%%%%%%%observation%%%%%%%%%%%%%%%%%%%%

%%%%%%%%%%%%%%%%%%%%analysis&results%%%%%%%%%%%%%%%%%%%%
\section{Determination of Polarization and Selection of Intrinsically Polarized Stars}
\subsection{Data Analysis and Determination of Polarization}
For the processed images, we carried out point spread function (PSF) photometry and aperture correction
with the IRAF\footnote{Image Reduction and Analysis Facility distributed by the National Optical Astronomy Observatory 
(NOAO), operated by the Association of Universities for Research in Astronomy, Inc. (AURAI) 
and under cooperative agreement with the National Science Foundataion (NSF).} tasks: $daofind$, $phot$, $psf$, and $allstar$.
After combining dithered images, the area we analyzed is 17.''2 $\times$ 17''.2 or 0.7\,pc $\times$ 0.7\,pc region.
We use stars whose positions are matched within one pixel in images for four waveplate angles. 
Then intensities in four images for each star are used to calculate the Stokes parameters $I$, $Q$, $U$, 
the degree of polarization $P$, and its position angle $\theta$ using the following equations:
$I = (I_0+I_{45}+I_{22.5}+I_{67.5}) / 2,
Q = I_0-I_{45},
U = I_{22.5}-I_{67.5}, 
P         = \sqrt{(Q/I) ^2+(U/I)^2}$, and $
\theta   = \frac{1}{2}\mathrm{arctan}(U/Q)$, 
where $I_{\rm x}$ is the intensity with the half wave plate oriented at $x$ deg.  
The errors $\delta P$ and $\delta \theta$ are calculated from the error propagation of photometric errors derived by IRAF. 
We normalized position angle $\theta$ 
by using 13 sources whose $\theta$ was determined with an accuracy of $<$20\,deg by \cite{ott99}.
We remove the bias of $P$ with the equation $P_{{\rm db}} = \sqrt{ P^2 - (\delta P)^2 }$ (\citealt{war74}),
where $P_{{\rm db}}$ is the debiased degree of polarization and $\delta P$ is the error of $P$. 
We calibrated the $K_S$-band magnitude of the stars 
using 61 bright ($m_{K_S}<14$) stars in the point source catalogue by \cite{sch10}.

\begin{figure}
\centering
\includegraphics[width=10cm,clip]{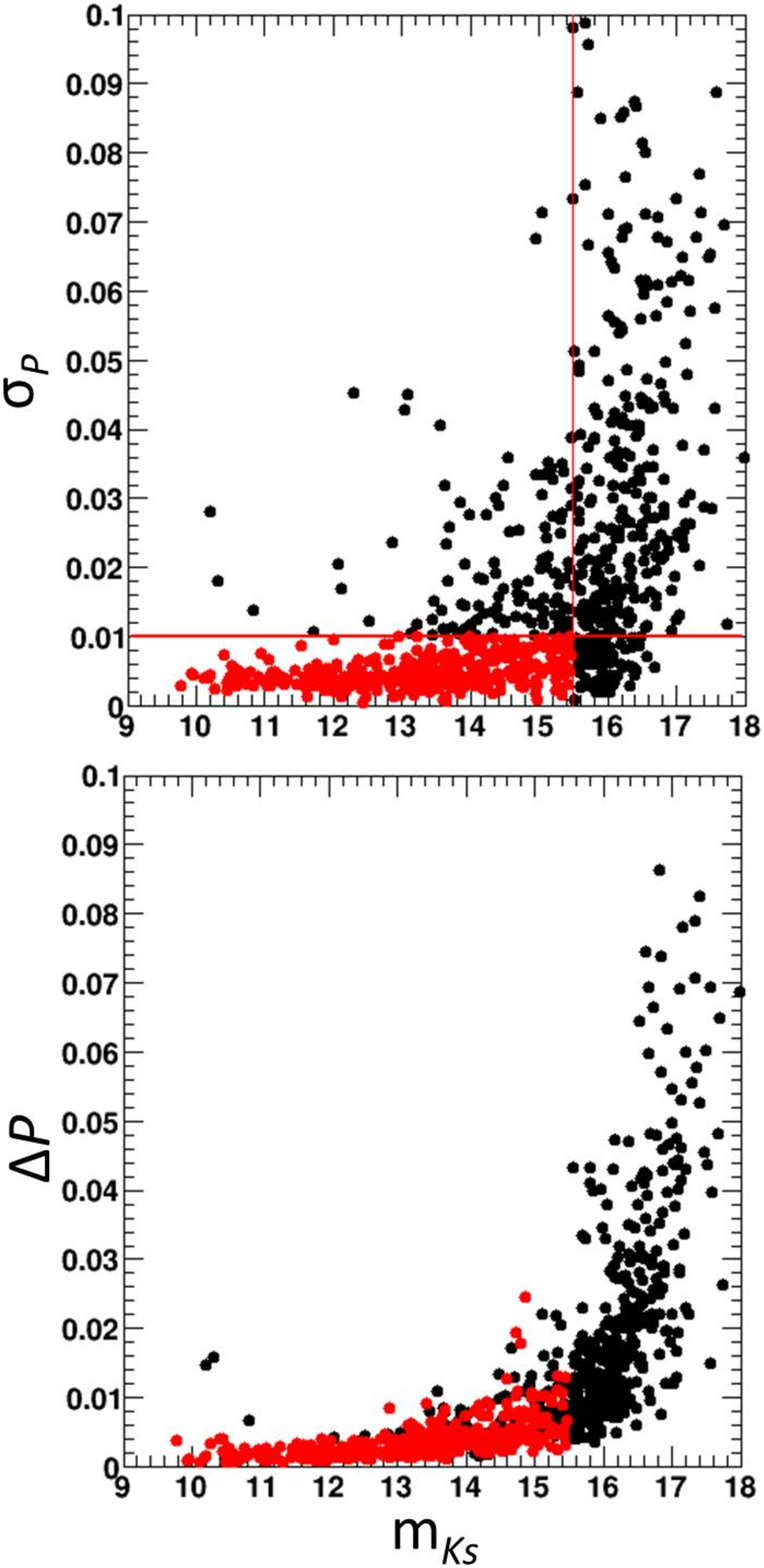}
\caption{Scatterplots of $\sigma_P$ vs. $m_{K_S}$ ({\it top}) and $\Delta P$ vs. $m_{K_S}$ ({\it bottom}), where  
$\sigma_P$ represents the standard deviation of polarization of each object during three nights (see text) 
and $\Delta P$ is polarimetric error obtained by IRAF/DAOPHOT photometry. 
Stars with $\sigma_P<0.01$ and $m_{K_S}<15.5$ are defined as ``good-reproducibility stars'' (red plot), 
and used in our analysis.}
\label{fig:mag_perr}
\end{figure}

Since the AO guide star is $\simeq$20$''$ far from the center of our observational field 
and the field of view is rather large, it can be seen in the raw image that the profile of PSF varies spatially, 
and in general it might be better to use a variable PSF for photometry.
However, in our field, there is an insufficient number of isolated stars 
to reliably compute variability of the PSF and calculate an aperture correction across the field of view.
We thus examine whether photometry with a variable PSF works better than a non-variable PSF by counting ``good-reproducibility stars''.
Here the ``good-reproducibility star'' is defined as a star whose polarization has a standard deviation
$\sigma_P = \sqrt{\sigma _{Q/I}^2 + \sigma _{U/I}^2}$ less than 0.01,
where $\sigma _{Q/I}$ and $\sigma _{U/I}$ are calculated from the three night data sets (Figure \ref{fig:mag_perr}). 
The numbers of the ``good-reproducibility stars'' are approximately 300 for both of the first-order-variable and non-variable PSF photometry.
We therefore adopt the non-variable PSF photometry.
Here the Moffat function\footnote{We use builtin function called ``moffat15'', 
which is an elliptical Moffat function with a beta parameter of 1.5.} 
is used to fit the PSFs, and after the photometry, we carried out an aperture correction.  
The number of the ``good-reproducibility stars'' with $m_{K_S} < 15.5$ is 319, 
and their errors are $\Delta P \approx 0.01$ (Figure \ref{fig:mag_perr}). 
Through this process, we dropped $\sim$10\,\% of unsaturated, bright stars in the range of $10<m_{K_S}<13$.

\cite{ott99} observed nearly the same region as ours and obtained polarization of 41 sources. 
We have measured polarization for 24 out of the 41 sources; 
other sources observed by \cite{ott99} are saturated in our data, or outside our observational field. 
To compare these two data sets, 
we calculate the flux-weighted average of $P$ and $\theta$ of the common 24 stars, 
resulting in 4.9\,\% $\pm$ 0.2\,\% and $24^{\circ} \pm 1^{\circ}$ for our sources (statistical error only), 
whereas 5.1\,\% $\pm$ 1.9\,\% and $34^{\circ} \pm 20^{\circ}$ for the \cite{ott99} data.
Thus, the average degree of polarization $P$ is consistent between the two data sets. 
Furthermore, although the results of \cite{ott99} have larger errors, 
our results of each star are in general consistent with theirs. 
For all of the 319 sources, we obtain the flux-weighted average of $P = 5.0\,\% \pm 0.1\,\%$ and $\theta = 23^{\circ} \pm 1^{\circ}$. 
We also obtain the average of $P = 5.2\,\% \pm 0.9\,\%$ and $\theta = 23^{\circ} \pm 6^{\circ}$, 
which are derived from the mean and the scatter in $Q/I$ - $U/I$ plane (see Section 3.2). 
These values are consistent with the previous studies of 
$P = 4.1\,\% \pm 0.6\,\%$ and $\theta = 30^{\circ} \pm 10^{\circ}$ for all sources within 20'' $\times$ 20'' \citep{ott99} and 
$P = 4.6\,\% \pm 0.6\,\%$ and $\theta = 26^{\circ} \pm 6^{\circ}$  for some parts of 15'' $\times$ 15''; \citep{buc11}.
Although \cite{buc11} find higher degrees of polarization and more north-south position angles appear 
in the east edge of their field of view, we do not find such a tendency.
To examine interstellar polarization toward the central parsec,
we exclude intrinsically polarized stars ($>3\,\sigma$) and calculate the flux-weighted average of $P$ and $\theta$. 
The results are $P$=4.9\,\% $\pm$ 0.1\,\% and $\theta=$23$^{\circ}$.6 $\pm$ 0$^{\circ}$.2 for 307 sources.
The position angle reflects the angle of the Galactic disk \citep[$\sim$31.4\,deg;][]{rei04}.

\subsection{Selecting Intrinsically Polarized Stars}
\begin{figure}
\centering
\includegraphics[width=12cm,clip]{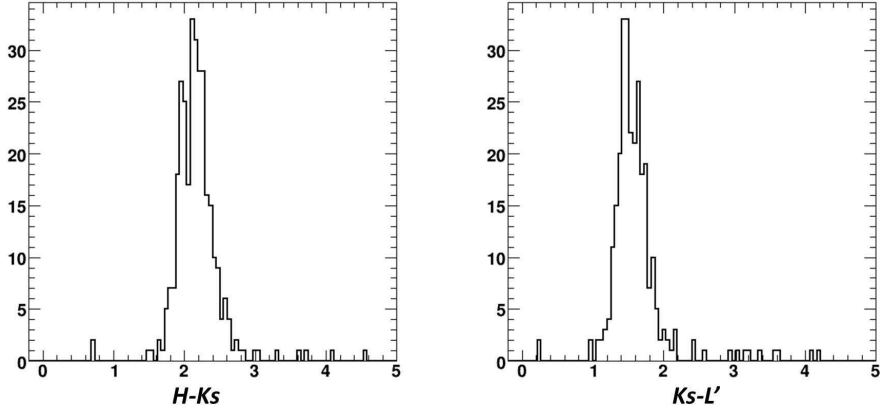}
\caption{Histograms of $H-K_S$ ({\it left}) and $K_S-L'$ ({\it right}) for ``good-reproducibility stars'', 
for which we use the $H$- and $L'$-band data from \cite{sch10}.
}
\label{fig:two_color_histo}
\end{figure}

To identify intrinsically polarized stars, we have to remove foreground stars first.
\citet{buc09} reported 58 foreground stars from near-infrared, intermediate-band imaging with AO 
in the field of view of $\sim$40''$\times$40''.
Only two stars (their source B\,275 and B\,458), whose $K$-band extinction $A_K$ are 1.0 and 1.9, respectively, 
are included in our list of the ``good-reproducibility stars''.
The observed color of B\,275 is small ($H-K_S=0.70, K_S-L'=0.22$) and therefore it is certainly a foreground star.
B\,458 has similar $HK_SL'$ colors to the Galactic center sources ($H-K_S=2.12, K_S-L'=1.43$),
and it is thus likely to be in the Galactic center.
We treat 318 stars excluding B\,275 in our analysis as stars in the central parsec.
This is also confirmed from the position of $H-K_S$ and $K_S-L'$ histogram 
\citep[using the $H$-, the $K_S$- and the $L'$-band data from][]{sch10}. 
In Figure \ref{fig:two_color_histo}, B\,275 has remarkably small values in both $H-K_S$ and $K_S-L'$. 
We also find that other 318 data are concentrated in the position of $H-K_S\sim2$ and $K_S-L'\sim1.5$. 
This color indicates they are sources in the central parsec \citep[see Figure 4 in][]{sch10}.

\begin{figure}
\centering
\includegraphics[width=12cm, clip]{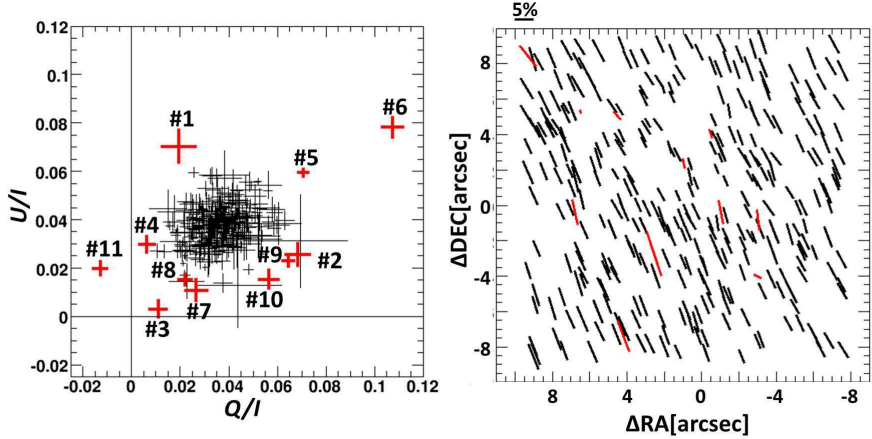}
\caption{{\it left}: A $Q/I-U/I$ plane - a scatterplot of $U/I$ vs. $Q/I$ for ``good-reproducibility stars'', 
excluding a foreground star (B\,275). 
The colors of data points represent the significance of intrinsic polarization of sources. 
Black cross represents $<3\,\sigma$ and red cross is $>3\,\sigma$. 
Here, $\sigma$ is the rms of photometric error and spread of data points. 
Number in this figure (\#1, \#2, and so on) corresponds to the column 1 in Table \ref{tab:fea}. 
{\it right}: A polarization vector map for ``good-reproducibility stars'', excluding the foreground star. 
The length and the angle of bars represent $P$ and $\theta$, respectively. 
The colors of bars have the same meaning as left figure.}
\label{fig:qu_plane_vector}
\end{figure}

\begin{figure}
\centering
\includegraphics[width=12cm, clip]{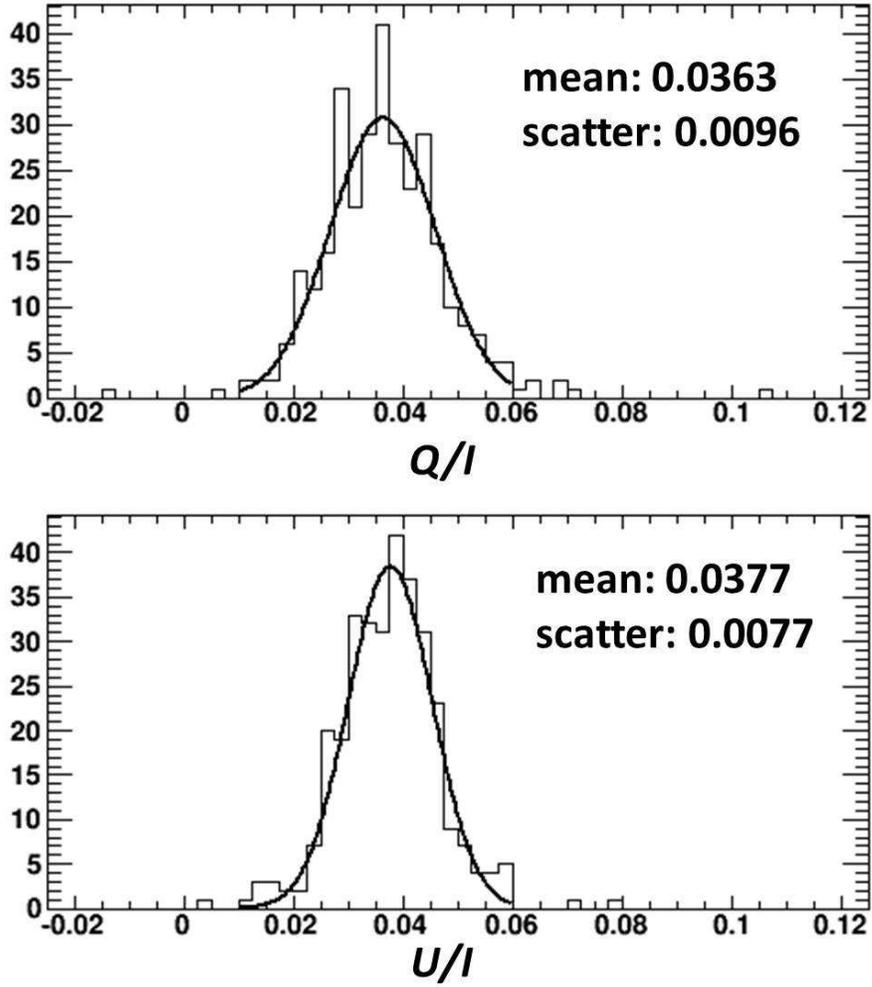}
\caption{Histograms of $Q/I$ ({\it top}) and $U/I$ ({\it bottom}) for ``good-reproducibility stars'', excluding a foreground star. 
Black solid curves are fitted with Gaussian functions.}
\label{fig:qu_hist}
\end{figure}

The left panel of Figure \ref{fig:qu_plane_vector} is 
a $Q/I-U/I$ plane presentation of the polarization of all the 318 stars, 
and the right panel is a polarization-vector map. 
In the $Q/I-U/I$ plane, the vast majority of stars are concentrated in a well defined region, detached from the origin. 
This detachment reflects interstellar polarization, 
which originates from the dichroic extinction by aligned dust grains along the line of sight.
The spread of the concentrated stars in the $Q/I - U/I$ diagram is estimated 
by fitting with Gaussian functions to remove the effect of outliers, 
and the standard deviations are 0.0096 in $Q/I$ and 0.0077 in $U/I$ (Figure \ref{fig:qu_hist}). 
The spread can be attributed to both uncertainties in measurement and real variation of interstellar polarization.

\begin{figure}
\centering
\includegraphics[width=12cm, clip]{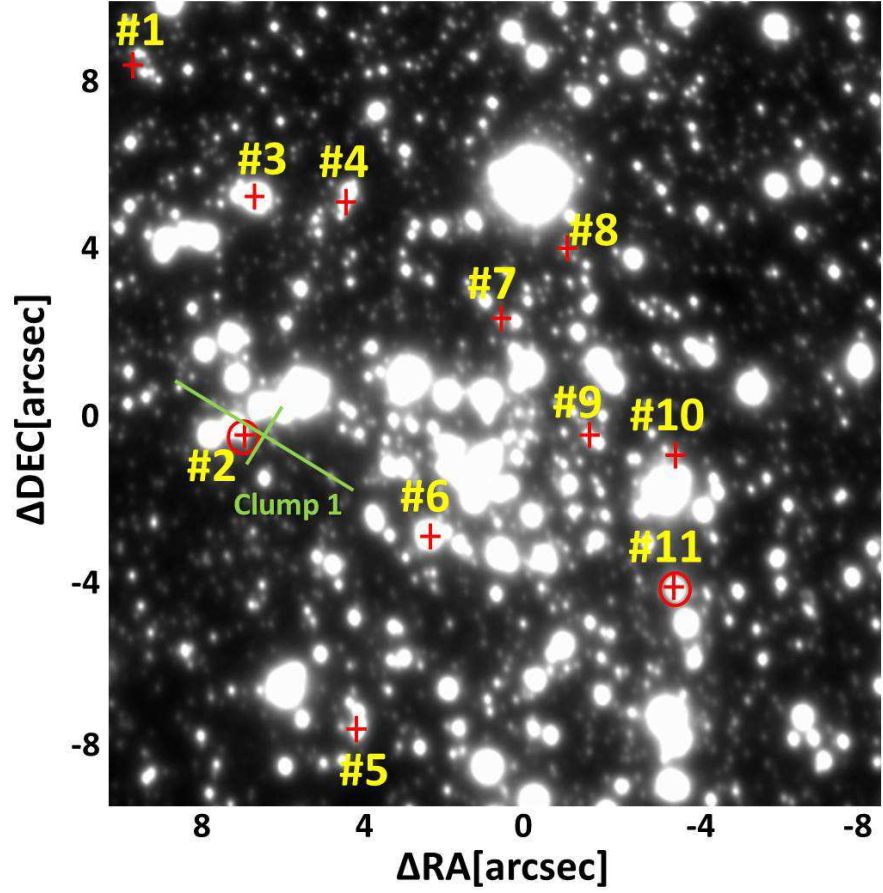}
\caption{The spatial distribution of intrinsically polarized stars.
Red cross represents a $>3\,\sigma$ polarized object and circled ones (\#2 and \#11) are good YSO candidates (see Discussion).
Green cross represents the position of SiO (5-4) Clump 1,  
and the length of cross corresponds to the spatial resolution of $2''.61 \times 0''.97$ \citep{yus13}.}
\label{fig:spatial_distribution}
\end{figure}

\begin{figure}
\centering
\includegraphics[width=12cm, clip]{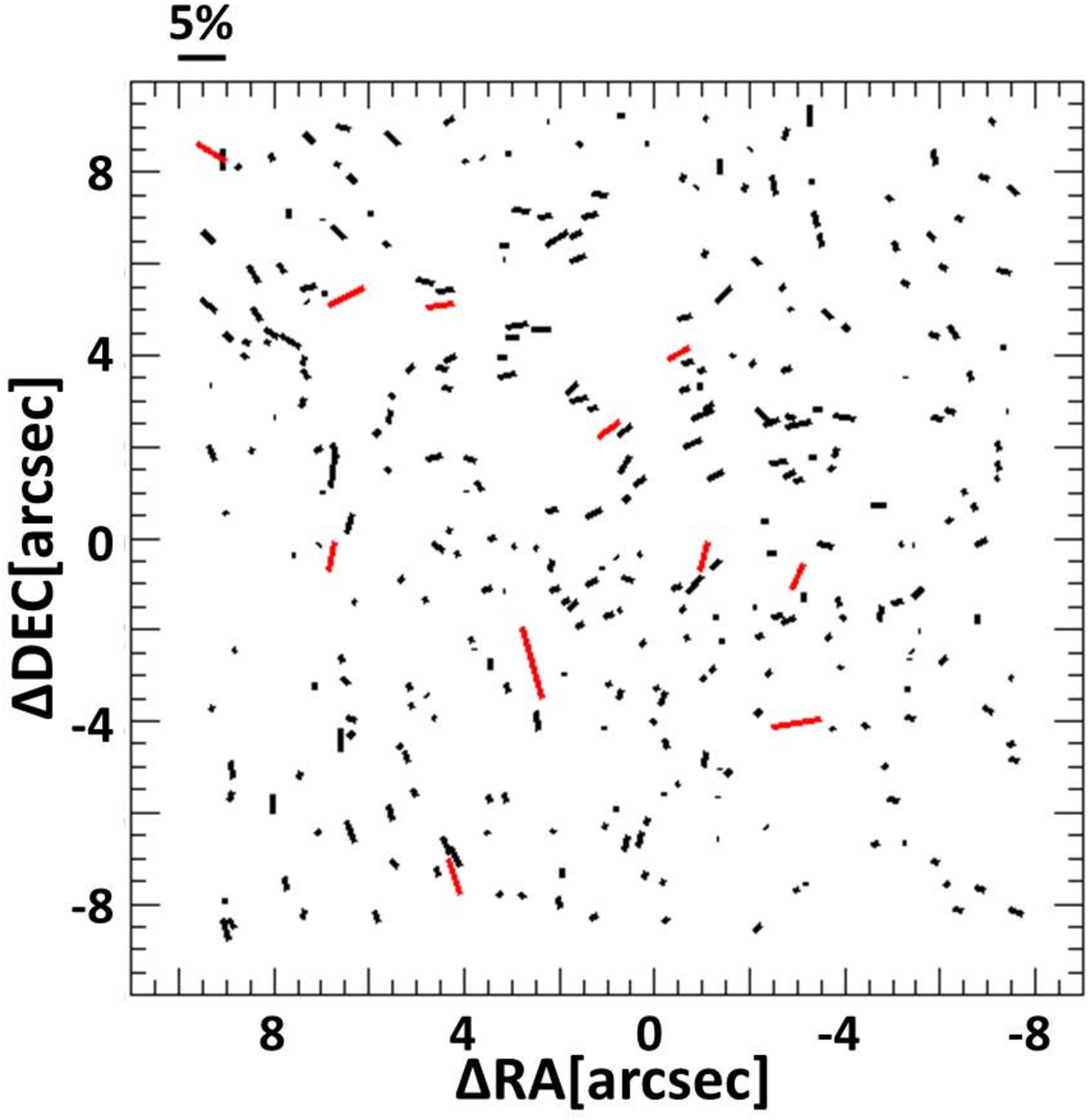}
\caption{Intrinsic polarization vector map. 
We calculate intrinsic polarization by subtracting the average of interstellar polarization in this field 
$(Q_{{\rm peak}}, U_{{\rm peak}})=(0.0363, 0.0377)$ from Stokes parameters $(Q, U)$ of each star. 
Red and black lines represent stars whose intrinsic polarization is detected as more than 3\,$\sigma$ and less than 3\,$\sigma$, respectively.}
\label{fig:int_vector}
\end{figure}

Since the spread is fairly well represented by a Gaussian function, 
and several stars deviate by amounts not explained by the errors and the variation in interstellar polarization, 
we can classify these stars as intrinsically polarized stars.  
In selecting them, we calculated the quadratic sum $\sigma$ of the polarimetric error of each star 
and the standard deviation of the Gaussian fitting. 
Here the intrinsically polarized stars are defined as stars 
which are apart from the peak of $Q/I$ and $U/I$ by $>3\,\sigma$ , and are listed in Table \ref{tab:fea}.
The spatial distribution of them is shown in Figure \ref{fig:spatial_distribution}. 
Figure \ref{fig:int_vector} shows the intrinsic polarization vector map.

\begin{figure}
\centering
\includegraphics[width=12cm, clip]{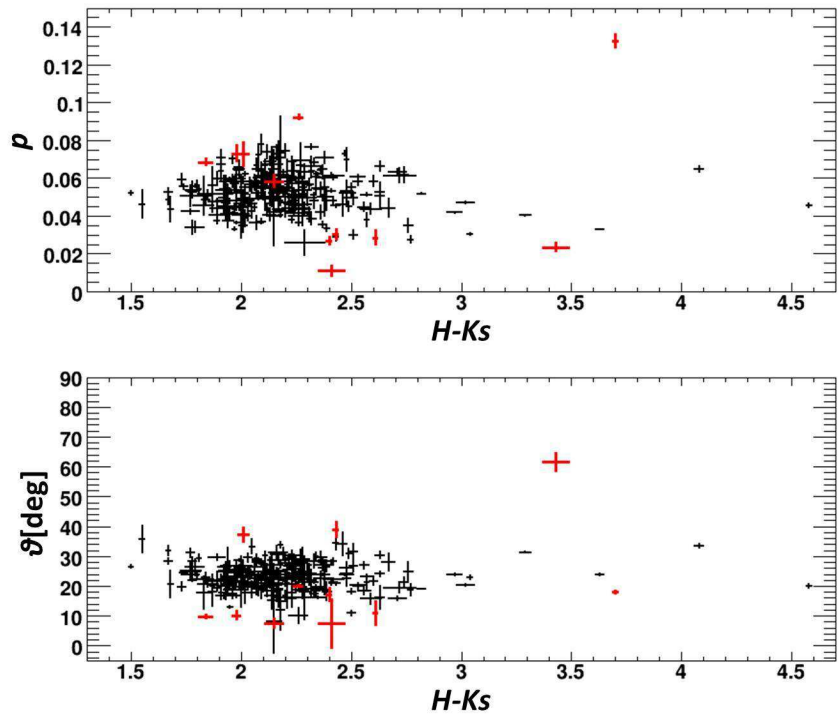}
\caption{Scatterplots of $P$ vs. $H-K_S$ ({\it top}) and $\theta$ vs. $H-K_S$ ({\it bottom}).
Red plot is a $>3\,\sigma$ polarized object.}
\label{fig:pol_color}
\end{figure}

The variation in the interstellar polarization in our stars is estimated in the following way.
We draw an $H-K_S$ histogram of the stars with m$_{K_S}<15.5$ using the catalogue of \cite{sch10} and fit the histogram with a Gaussian function. 
The resultant $H-K_S$ variation is $\sim$0.23\,mag. 
Since K-type giants are dominant with m$_{K_S}<15.5$ in the central parsec and 
very little $H-K_S$ variation is expected in their intrinsic colors, 
we assume that the variation in $H-K_S$ is entirely due to the interstellar extinction. 
Concerning the relation between interstellar polarization and interstellar extinction, 
\cite{hat13} derive the polarization efficiency toward the central 300\,pc, $P_{K_S}/E(H-K_S) = $2.4\,\%/mag.
$E(H-K_S)$ represents color excess of $H-K_S$, and on the assumption that K-type giants are dominant in the central parsec, 
we regard the variation of $H-K_S$ as that of $E(H-K_S)$.
Therefore, we estimate the variation of $K_S$-band interstellar polarization is $\sim$0.55\,\%. 
This value is smaller than the spread of data points in $Q/I - U/I$ diagram ($\sim0.87\,\%$; Figure \ref{fig:qu_plane_vector}).
Since the spread in $Q/I - U/I$ diagram contains the variation of interstellar polarization, 
our selection of intrinsically polarized stars is conservative.  
Toward the central parsec, there seems to be weak or no dependence of interstellar polarization 
on the amount of the interstellar extinction (Figure \ref{fig:pol_color}).
Similar results are reported by \citet[see their Fig. 16]{buc11}.
This might be related to the higher magnetic field strength of the random component 
compared to that of uniform component between the central parsec and us \citep{hat13}.

%%%%%%%%%%%%%%%%%%%%analysis&results%%%%%%%%%%%%%%%%%%%%

%%%%%%%%%%%%%%%%%%%%discussioin%%%%%%%%%%%%%%%%%%%%
\section{Discussion}
\subsection{Colors of Intrinsically Polarzied Stars and Their Implication for Star Formation in the Central Parsec}
For YSO identification, infrared color-color diagrams are useful
in assessing whether any particular star has excess emission.  
YSOs often exhibit infrared excesses due to thermal emission 
from the circumstellar disks and envelopes \citep[e.g.,][]{hil92,ish98,fue02}. 
Color-color diagrams can be used to distinguish between normal stellar colors that are reddened by intervening interstellar dust, 
and a contribution that is due to circumstellar emission. 
In particular, the $L$ band is sensitive to infrared excess produced by circumstellar disks 
and likely the optimum wavelength for detecting infrared excesses from circumstellar disks with ground-based telescopes \citep{lad00}. 

\begin{figure}
\centering
\includegraphics[width=10cm, clip]{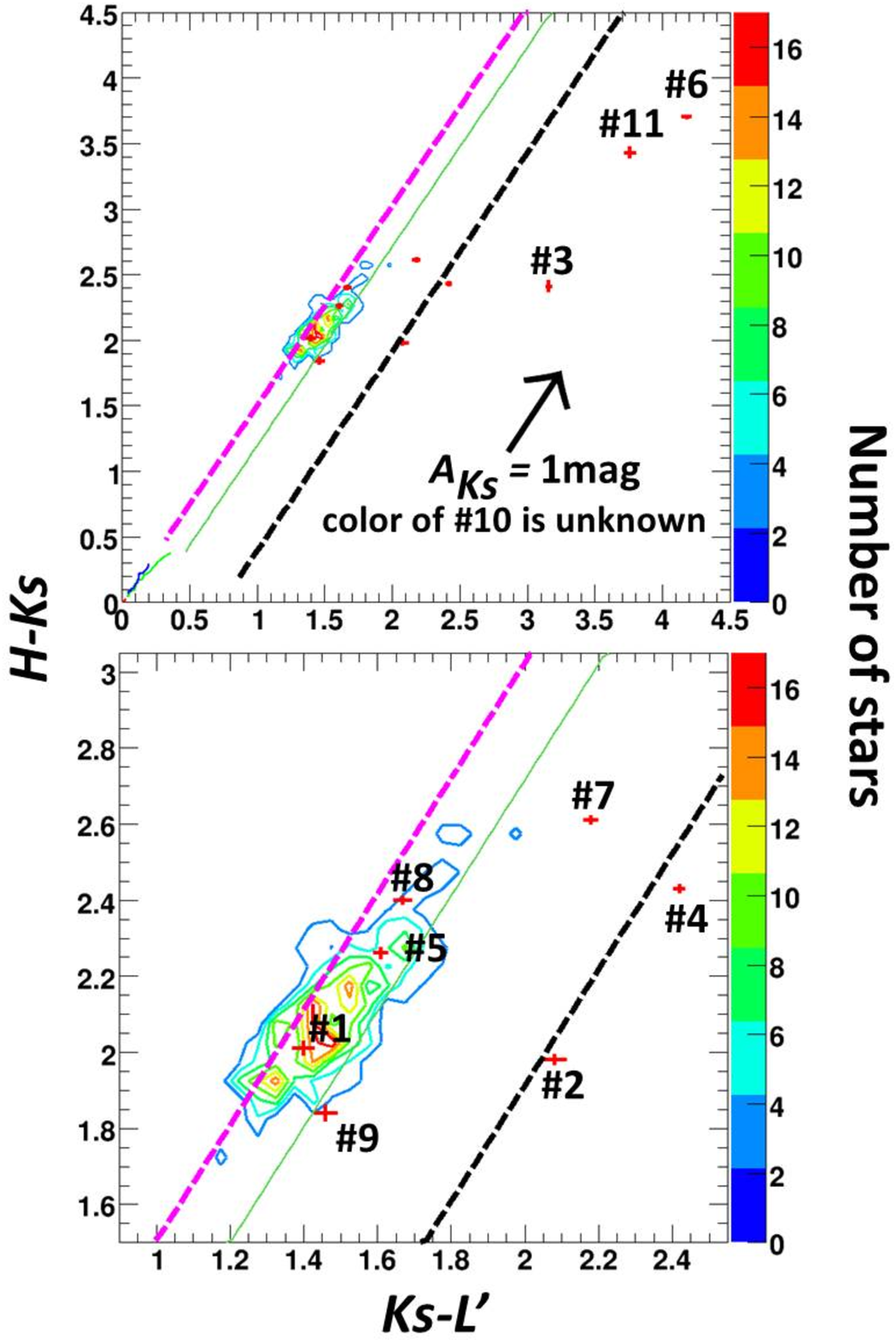}
\caption{$HK_SL'$ color-color diagram. 
The contour is drawn with stars in the field of our observations, 
whoes ${K_S}$-band magnitudes are less than 15.5\,mag and $L'$-band magnitudes are obtained in \citep{sch10}. 
The red cross represents a $>3\,\sigma$ polarized object.
Two dashed lines represent the extended loci along reddening vector \citep{sch10}. 
The magenta dashed line is for OBA-type main sequence + GKM-type giant \citep{bes88} 
and the black dashed line is for T Tauri stars \citep{mey97}. 
Solid lines in the lower left corner of this figure represent the colors of OBA dwarfs (red), FGKM dwarfs (green), 
GKM giants (blue), OBA supergiants (yellow) and FGKM supergiants (magenta), respectively \citep{bes88}. 
The contour of color of all stars in the observational field are superposed 
and green thin solid line is 1\,$\sigma$ of dispersion 
when we fit the histogram of color of stars with a Gaussian function along the vertical direction to reddening vector. 
Lower panel is zoom-in view. Note that one 3\,$\sigma$ source is not plotted 
because the $L'$-band magnitude is not in the catalogue of \citep{sch10}.}
\label{fig:two_color}
\end{figure}

\begin{figure}
\centering
\includegraphics[width=12cm, clip]{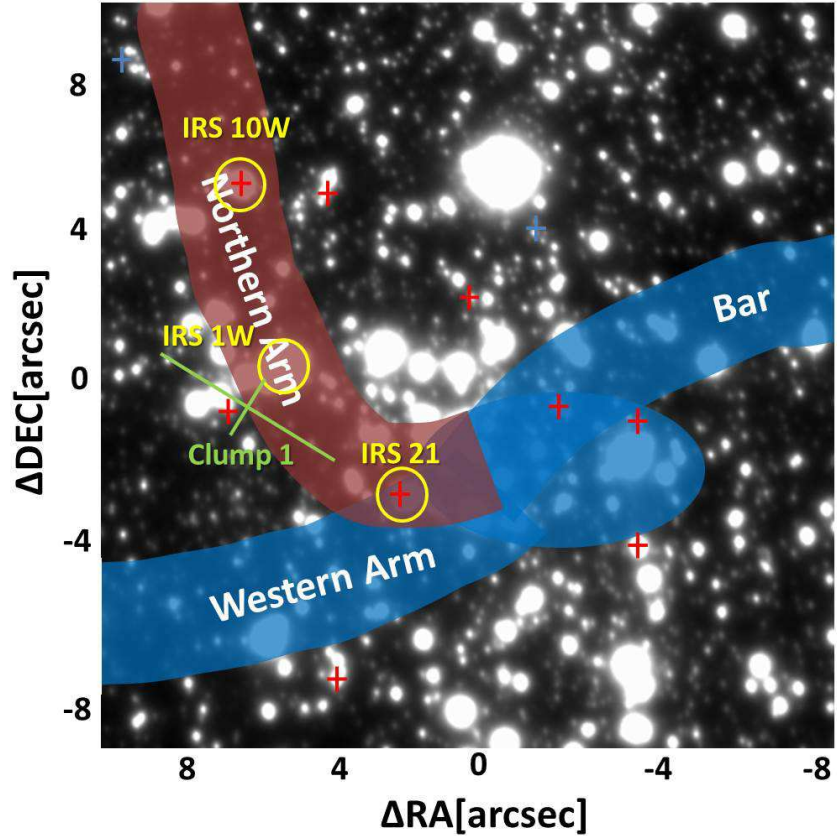}
\caption{The schematic view of the minispiral is superimposed 
on the $K_S$-band image and the spatial distribution of intrinsically polarized stars. 
Red cross has infrared excess in Figure \ref{fig:two_color}, and blue cross does not.
Green cross represents the position of SiO (5-4) Clump 1,  
and the length of cross corresponds to the spatial resolution of $2''.61 \times 0''.97$ \citep{yus13}.
Bow shock sources along the Northern Arm studied by \cite{tan02,tan05} are indicated by yellow circles.}
\label{fig:comparing}
\end{figure}

In Figure \ref{fig:two_color}, we draw an $H-K_S$ versus $K_S-L'$ color-color diagram of intrinsically polarized stars 
on the basis of the infrared three-bands catalogue of the central parsec \citep{sch10}. 
The intrinsically polarized stars are rather scattered in Figure \ref{fig:two_color}, 
in contrast to the concentration of the majority of stars in the field 
along the reddening locus from main sequence and giant stars.  
Seven of the intrinsically polarized stars are redder than the 1-$\sigma$ locus (see Figure \ref{fig:two_color}), 
and they are shown in Figure \ref{fig:comparing} as red crosses.
These seven stars probably have a circumstellar envelope which is seen nearly edge on \citep[see Figure 9 in][]{rob06}. 
In particular, Stars \#6 (IRS\,21), \#11 (IRS\,2L) , and \#3 (IRS\,10W) 
are even much redder than the reddest T Tauri locus.  
IRS\,21 and IRS\,10W were identified as near-infrared excess sources embedded in the minispiral and having a bow shock \citep{tan02, tan05}.
\citet{vie06} called them Northern Arm bow shock sources.  

\begin{figure}
\centering
\includegraphics[width=12cm, clip]{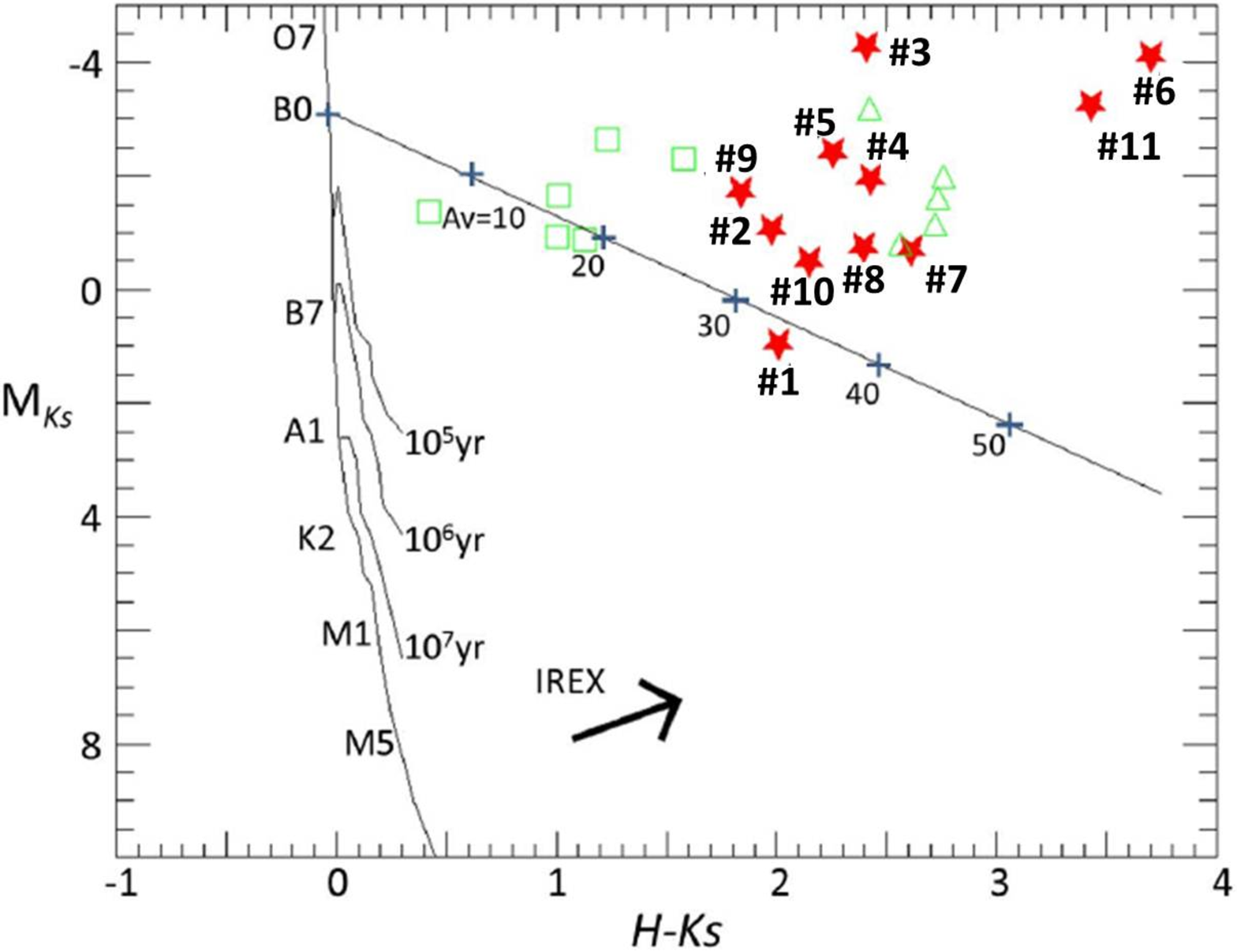}
\caption{Color-magnitude diagram (M$_{K_S}$ vs. $H-K_S$) of our YSO candidates and other massive YSO candidates. 
Superposed are the main sequence (leftmost curve), reddening line (diagonal line), 
isochrones for pre-main sequence stars (four full lines) and, infrared-excess arrow \citep[Figure 5 in][and references therein]{fau09}. 
Filled star marks represent our YSO candidates and the numbers corresponds to the numbers in \ref{tab:fea}. 
Green symbols represent massive YSO candidates from other studies: squares are from \cite{han97} and triangles are from \cite{blu01}.}
\label{fig:faustini}
\end{figure}

Additional indications of the intrinsically polarized stars may be derived 
from the $K_S$ versus $H-K_S$ color-magnitude diagram (CMD, Figure \ref{fig:faustini}).
The intrinsically polarized stars occupy a region of 
massive YSO candidates and OB stars reddened by $A_V=30-50$\,mag in Figure \ref{fig:faustini}.  
In fact, the $K-L$ versus $K$ CMD in \citet{cle01} has 
the Northern Arm bow shock sources IRS\,1W and IRS\,21, 
which were classified as YSO candidates at that time, in the rightmost region with the reddest colors 
in the similar manner to Figure \ref{fig:faustini}.  
Then in the $K-L$ versus $K$ CMD \citep{cle01}
are late-type Wolf-Rayet (WR) stars such as WC9 in the middle,
and AGB stars and Ofpe/WN9 stars in the leftmost region. 
If some of the intrinsically polarized stars are genuine YSOs, 
they are very young ($\lesssim10^{4-5}$\,yr) and massive ($\gtrsim6-8$M\,$_{{\sun}}$).  
The presence of such young stars would indicate
that these stars were formed in the central parsec, near the SMBH. 

According to the ``inspiraling star cluster'' scenario, 
the required mass of a stellar cluster to infall into the central parsec from 30\,pc 
within $\sim 10^5$\,yr is $\sim 6\times 10^7$M\,$_{{\sun}}$ \citep{ger01, gur05}, which is unreasonably massive.
Even if we set the initial position of a stellar cluster to be 10\,pc, 
the required mass is $\sim 6\times 10^6$M\,$_{{\sun}}$, as massive as Omega Centauri (NGC 5139). 
Considering the star clusters in the Galactic center (Arches, Quintuplet, and Central cluster), 
the typical mass of star cluster seems to be $\sim10^4$M\,$_{{\sun}}$
and it is not reasonable to assume $\sim 6\times 10^6$M\,$_{{\sun}}$. 
Although the circumnuclear disk (CND), which is a ringlike structure of molecular gas and dust, 
surrounding Sgr A* \citep[see Figure 13 and Table 2 in][]{chr05}, 
is the most massive and densest molecular cloud within the central 10\,parsec, 
the total gas mass of the CND is $\sim10^6$M\,$_{{\sun}}$, 
and it is not sufficient for the formation of a star cluster like Omega Centauri.
Therefore, such young stars found in the central parsec must have been formed near the SMBH and 
we can reject the ``inspiraling star cluster'' scenario if some of the intrinsically polarized stars we found are genuine YSOs.
Recent studies suggest that young massive stars in the central parsec were formed in situ $\simeq6\pm$2\,Myr ago 
in a burst of star formation whose duration is less than 2\,Myr \citep{pau06,bar09}. 
Since the estimated ages of the putative YSOs are significantly smaller than 6\,Myr, 
they were possibly formed in the remnant gas of the gaseous disk. 
As another possibility, small-scale star formations may have been occurring intermittently in the central parsec. 
\cite{chr05} revealed that there are 26 dense molecular gas cores in the CND. 
The enhanced core densities and masses may explain the formation of massive young stars.  

\subsection{Classification of Intrinsically Polarized Stars}
Some of the intrinsically polarized stars have been spectroscopically observed \citep[e.g.,][]{pau06}, 
and furthermore, most of them have been classified into late and early-type stars 
on the basis of photometry with narrow-band filters around $2.3\,\mu$m \citep{buc09}.  
\cite{buc09} classified stars in the central parsec into late- and early-type stars 
on the basis of characteristic CO bandhead absorption
which late-type stars show in the longer wavelength part of the $K$ band. 
They demonstrate that the CO bandhead depths of stars previously classified spectroscopically 
fall into separate regions although a few (less than 5\%) early-type stars are classified as late-type and vice versa.
In our sample, Star\,\#5 could be one example of contaminations, which is classified as late-type in \cite{buc09},
but it is classified as a hot ``He star'' using $KLM$ two-color diagram in \cite{vie05}.

Here, we classify the intrinsically polarized stars into three groups: bow shock sources in the Northern Arm (\#2, \#3, \#4, and \#6), 
other early-type stars (\#9 and \#11), and late-type stars (\#1, \#5, \#7, \#8, and \#10). 
We discuss each group by referring to the literature.

\subsubsection{Bow Shock Sources in the Northern Arm (\#2, \#3, \#4, \#6)}
The Galactic center sources IRS\,1W, 5, 8, 10W (Star\,\# 6), and 21 (Star\,\# 3) are
all WR and O type stars producing bow shocks with strong winds, plowing through the
ambient gas and dust of the Northern Arm, which is one stream of the minispiral \citep{tan05, geb06}. 
The positions of IRS\,1W, 10W, and 21 are shown in Figure \ref{fig:comparing} as yellow circles.
These sources have nearly featureless near-infrared spectra with infrared excess and 
often exhibit polarization that cannot be accounted for 
by interstellar polarization \citep{kra95,ott99}. 
Due to these features, they were first believed to be prime YSO candidates \citep[e.g.,][]{cle01}. 
However, recent studies show that they are early-type stars whose stellar winds generate bow shocks in the Northern Arm. 
Therefore the infrared excess is due to re-radiation from heated dust grains and 
the polarization comes from scattering of the stellar light by non-spherically distributed dust \citep{tan05}.  

Three of the bow shock sources reported by \cite{tan02,tan05} are in the current observation field of view.  
IRS\,10W (Star\,\#3) and IRS\,21 (Star\,\#6) exhibit intrinsic polarization in our observations.
IRS\,1W is not included in our list of the intrinsically polarized stars 
because we were able to obtain polarization of IRS\,1W only during the first night
due to saturation (m$_{K_S}=9.27$ according to the first night observation).
If we use the first night data, IRS\,1W exhibits intrinsic polarization: 
$P_{{\rm int}}=8.27\%$ and $\theta_{{\rm int}}=110.9^{\circ}$.

In the Northern Arm, Stars\,\#2 and \#4 are also intrinsically polarized.
These stars are classified as early-type stars in \cite{buc09}, 
and Star\,\#2 is classified as WC9 and Star\,\#4 is Ofpe/WN9 in \cite{pau06}.
Star\,\#2 might be a good YSO candidate as we mention below, but it certainly seems to be accompanied with outflow activity.
Though WR stars exhibit intrinsic polarization due to scattering of free electrons
in the circumstellar envelope ejected as stellar wind,
the degree of polarization $P$ is usually less than 1\,\% \citep[e.g.,][]{rob89}.
However, \#2 and \#4 show intrinsic polarizations of more than 3\%.
This high polarizations could be caused by the interaction between stellar wind and the minispiral, 
although a few WR stars do exhibit a large intrinsic polarization \citep[3-4\,\%, e.g.,][]{vil06}.

Star\,\#2 is located in one of the 11 SiO (5-4) clumps of molecular gas detected by \cite{yus13}.
They interpreted the SiO sources as YSO outflows, and identified a YSO candidate (source 526311) which drives the outflow,
in the {\it SPITZER} IRAC point source catalogue \citep{ram08}.
IRAC source 526311 is assigned the magnitudes of the near-infrared 
$JHK_S$ ($m_J=13.304$, $m_H=9.439$, and $m_{K_S}=7.898$) and the position of 2MASS. 
The position of Star\,\#2 agrees with the peak of Clump 1 very well, 
and thus Star\,\#2 could be the driving YSO of Clump 1.

\cite{tan05} argue that the observed bow shock structures are generated
by the interaction of stars rapidly moving through the Northern Arm flow, 
and that the bow shocks heat and perturb the dust grains associated with the Northern Arm.
In this model, the intrinsic polarization angle of the source should be perpendicular to
the position angle of bow shock structure (Table 3 in \citealt{tan05}) 
because the position angle caused by dust scattering is perpendicular to 
the direction from dust to the illuminating source.  
The position angle of bow shock structure is calculated from 
the relative velocity vector of the source to the flow of the minispiral.   
We examine this relation in Table \ref{tab:pa}, where we calculate the position angles of bow shocks
using the velocities of sources in \cite{pau06} and \cite{tan05} 
and the flow of the Northern Arm in \cite{pau04}. 

The predicted position angles of bow shocks, however, do not necessarily 
agree with the direction perpendicular to the position angles of intrinsic polarization $\theta_{\rm int}$; 
Stars\,\#3, \#4, IRS\,1W are nearly perpendicular, but others are not.     
The discrepancy between these angles may indicate that the mechanism for generating bow shocks is not so simple.
The difference was already noticed for the bow shock structure and the prediction 
based on the relative velocity in IRS\,10W by \cite{tan05}.  
Also, note that \cite{tan05} did not detect any azimuthal asymmetry in 
IRS\,21\footnote{However, \citet{buc11} have indicated that 
this source is not circular in projection after applying a Lucy-Richardson deconvolution.  
}, 
and they interpreted this as a face-on bow shock.
If this interpretation is correct and polarization is produced by scattering by surrounding dust, 
IRS\,21 should have no large intrinsic polarization.
On the contrary to this interpretation, past polarimetric studies and our observation show 
large intrinsic polarization.  
\cite{ott99} explain the polarization of IRS\,21 by emission
from hot dust aligned by the magnetic field inside the minispiral, 
referring to \cite{ait91,ait98}, who observed mid-infrared emission in the central parsec
and derived the magnetic field inside the minispiral.  
The relation between bow shock and polarization angle is still unknown.

In these bow shock sources, Star\,\#2 is a good YSO candidate.
This star has intrinsic polarization and infrared excess, and could be the counterpart of SiO (5-4) clump.
These features coincide with those of early-stage ($10^{4-5}$\,yr) YSOs. 

\subsubsection{Other Early-type Stars (\#9, \#11)}
Two other intrinsically polarized stars, \#9 and \#11, are classified as early-type stars by \citet{buc09} and \citet{cle01}, respectively.
Star\,\#11 was named IRS\,2L in \cite{vie06}, and is one of the enigmatic sources in the central parsec.
IRS\,2L is one of the sources whose $M$-band spectra were taken 
and the line-of-sight absorption to IRS\,2L was examined by \cite{mou09}.  
They used the IRS\,2L spectrum as a template for the foreground absorption
on the assumption that IRS\,2L is located close but not inside the minispiral and is not behind the circumnuclear-disk material. 

IRS\,2L is classified as a Be star by \cite{cle01}. 
It is known that Classical Be stars exhibit polarization
due to scattering of free electrons in a circumstellar gaseous non-spherical shell \citep{yud00}. 
However, the degree of polarization $P$ of classical Be stars is usually small ($0 \% < P < 1.5\%$ for 95\,\% of Be stars), 
and only 10 out of 495 Be stars exhibit $\sim$2\,\% \citep{yud01}. 
\citet{wat92} also noted that in a framework of the single-scattering approximation 
and geometrically thin disks it is difficult to obtain high levels of polarization above 2\,\%, 
and there is much evidence that circumstellar envelopes are optically thin \citep{yud01}.  
Furthermore, since self absorption of starlight in free-free and free-bound transition at their disk acts 
as a source of depolarization \citep{coy69}, $K_s$-band polarization declines to $\sim$1\,\% 
\citep[Pf limit (2.28\,$\mu$m) is in the $K_S$ band; see also figures of][]{mcd01}.
Therefore, the high intrinsic polarization of IRS\,2L is not likely due to the Be star features.
  
\cite{vie06} studied the infrared SED of IRS\,2L 
and find that IRS\,2L has a SED of ``typical luminous bow shock sources'' like IRS\,21 and IRS\,1W.  
Since IRS\,2L does not seem to be located inside the minispiral area \citep{mou09}, 
the red and featureless SED might indicate that this is a genuine YSO.  
We notice that other dusty sources such as IRS\,13N \citep{eck12} are still YSO candidates.  

Star\,\#9 is classified as an early-type star in \cite{buc09}
and O8-9.5 III/I in \cite{pau06}.  
This star is not located in the Northern Arm, 
where \cite{tan02, tan05} identified the dusty sources as not YSOs but as the bow shock sources, 
so its strong intrinsic polarization cannot be explained by the existence of a bow-shock.
Thus, although \cite{pau06} identified this star as 
a member of the clockwise rotating system whose age is $\sim 6 \pm 2$Myr, 
its high degree of polarization suggests the presence of circumstellar matter, 
and its age might not be so certain. 

In this catagory, we regard Star\,\#11 as a good YSO candidate.
This star, IRS\,2L, has intrinsic polarization and infrared excess.
If this star is not under the effect of the minispiral, this star could not be a bow-shock source but a YSO.

\subsubsection{Late-type Stars (\#1, \#5, \#7, \#8, \#10)}
Five of the intrinsically polarized stars (\#1, \#5, \#7, \#8 and \#10) are classified as late-type stars in \cite{buc09}.
In general, late-type stars such as red giants exhibit CO absorption, while early-type stars like OB main-sequence stars do not.
However, some YSOs which are in the stage of mass accretion exhibit CO absorption \citep{hof06}.
According to the study, CO absorption is most likely to be a sign of heavily accreting protostars 
with high mass accretion rates above $10^{-5}$M$_{\sun}$yr$^{-1}$.
As shown in the CMD (Figure \ref{fig:faustini}), 
our intrinsically polarized stars could be very young and 
be in such a heavy accretion stage.
We also note that Star\,\#5 is classified as a late-type star in \cite{buc09}, 
but has an $L-M$ color similar to He stars, 
which are slightly but systematically redder than expected for hot stars \citep{vie05}.  
This might indicate the presence of circumstellar material.   
However, YSOs exhibit only weak CO absorption, if at all, 
and it is questionable that YSOs with weak CO absorption are classified 
as late-type in \cite{buc09} by narrow-band photometry.  
Therefore, spectroscopic studies of such a star is necessary for a firm classification.

Other possible candidates which exhibit both intrinsic polarization and CO absorption 
include red supergiants and post-AGB stars.
Red supergiants show polarization up to several percent \citep[e.g.,][]{yud00}. 
However, in brightness, the five intrinsically polarized stars are unlikely to be red supergiants.
In the central parsec, red supergiants are brighter than $m_K<10.4$ \citep{kra95}.
Some post-AGB stars show high polarization originating due to scattering of the stellar light 
by dust grains in the circumstellar envelopes, which was ejected
at the final mass-loss phases \citep[e.g.,][]{gle05}.
However, \cite{muz10} calculated the expected number of post AGB star,
and found less than one post AGB star in the central parsec (for more details, see 7.3 in \cite{muz10}). 
This indicates that it is not reasonable to regard all of the five polarized stars as post AGB stars.

Thus, we cannot identify what are the five intrinsically polarized stars.  
Although interaction with the minispiral is possible,  
just like the case of the WR stars and bow shocks, 
most of them are not exactly in the minispiral.

%%%%%%%%%%%%%%%%%%%%discussioin%%%%%%%%%%%%%%%%%%%%

%%%%%%%%%%%%%%%%%%%%summary%%%%%%%%%%%%%%%%%%%%

\section{Summary}
Our high angular resolution polarimetry in the $K_S$ band 
have revealed intrinsically polarized stars in the central stellar cluster. 
After subtraction of the interstellar polarization component, 
11 intrinsically polarized stars ($>3\,\sigma$) have been selected.
Stars\,\#3 (IRS\,10W) and\#6 (IRS\,21) are bow shock sources in the Northern Arm, 
and Star\,\#4 can be a similar source.  
These are believed to be WR stars producing bow shocks
with strong winds, plowing through the ambient gas and dust, 
but further studies are necessary to explain the observed bow shock images and observed polarization. 
In the remaining eight stars, two sources (\#2 and \#11) are good YSO candidates 
because they are early type and have infrared excess.
Moreover, Star\,\#2 could be the counterpart of the SiO Clump 1, and Star\,\#11 does not exist inside the minispiral.  
Observations with higher angular resolution would be of interest.  
Also detailed spectroscopic identifications of 
the other five stars thought to have CO absorption and two early-type stars is necessary.

%%%%%%%%%%%%%%%%%%%%summary%%%%%%%%%%%%%%%%%%%%

\acknowledgments
We are grateful to R. M. Buchholz for kindly providing the narrow-band photometry data. 
This work was partly supported by the Grant-in-Aid for JSPS Fellows for young researchers (T.Y. and S.N.).
This work was also supported by KAKENHI, 
Grant-in-Aid for Research Activity Start-up 23840044,
Grant-in-Aid for Specially Promoted Research 22000005,
Grant-in-Aid for Young Scientists (A) 25707012, 
Grant-in-Aid Scientific Research (C) 21540240, 
the Global COE Program ``The Next Generation of Physics, Spun from
Universality and Emergence'', and Grants for Excellent Graduate Schools, from the Ministry of Education,
Culture, Sports, Science and Technology (MEXT) of Japan.
We especially thank the anonymous referee for constructive comments which have significantly improved the manuscript.

\newpage

\begin{landscape}
\begin{table}[tb]
\footnotesize
\caption{The features of intrinsically polarized stars ($>3\,\sigma$).}
\begin{center}
\begin{tabular}{cccccccc} \hline
ID &  $\Delta$RA $^a$ & $\Delta$DEC $^a$ & m$_{K_s}$ & ($Q/I$, $U/I$) & $p_{{\rm int}}$, $\theta_{{\rm int}}$ $^b$ & {\small ($H-K_s$, $K_s-L'$)} & identification/      \\ 
      & (arcsec)     & (arcsec)    & (mag)     &           & (\%, deg)                        &    (mag)        & classification \\ \hline   
1     & 9.59  & 8.56  & 15.48  & (0.020, 0.070)   & (3.6$\pm$0.7, 59$\pm$5)   & (2.01, 1.40)   & late$^d$   \\ 
2     & 6.91  & -0.38 & 13.48  & (0.069, 0.026)   & (3.4$\pm$0.5, 170$\pm$4)  & (1.98, 2.08)   & WC9?$^c$, early$^d$   \\
3$^h$ & 6.60  & 5.38  & 10.26  & (0.011, 0.003)  & (4.3$\pm$0.3, 117$\pm$2)  & (2.41, 3.16)   & IRS\,10W$^e$  \\
4     & 4.49  & 5.17  & 12.60  & (0.006, 0.030)  & (3.1$\pm$0.3, 97$\pm$3)   & (2.43, 2.42)   & Ofpe/WN9, IRS\,7E2$^c$            \\ 
5     & 4.17  & -7.49 & 12.12  & (0.071, 0.059)   & (4.1$\pm$0.2, 16$\pm$1)   & (2.26, 1.61)   & late$^d$, He\,\,star$^f$\\
6$^h$ & 2.43  & -2.76 & 10.42  & (0.107, 0.078)    & (8.2$\pm$0.4, 15$\pm$1)   & (3.70, 4.18)   & IRS\,21$^e$ \\
7     & 0.74  & 2.41  & 13.83  & (0.027, 0.011)   & (2.8$\pm$0.4, 125$\pm$4)  & (2.61, 2.18)   & late$^d$  \\
8     & -0.83 & 4.11  & 13.80  & (0.022, 0.015)   & (2.7$\pm$0.2, 119$\pm$3)  & (2.40, 1.67)   & late$^d$  \\
9     & -1.40 & -0.38 & 12.81  & (0.065, 0.023)   & (3.2$\pm$0.2, 166$\pm$2)  & (1.84, 1.46)   & O8-9.5 III/I, W10$^c$, early$^d$ \\  
10    & -3.51 & -0.80 & 14.05  & (0.057, 0.015)   & (3.0$\pm$0.4, 156$\pm$4)  & (2.15, -)      & late$^d$ \\ 
11    & -3.49 & -4.06 & 11.29  & (-0.013, 0.020)  & (5.2$\pm$0.3, 100$\pm$2)  & (3.43, 3.76)   & IRS\,2L$^e$, Be star$^g$  \\ \hline
\end{tabular}
\end{center}
{\footnotesize
$^a$ The origin is Sgr A* (RA = 17:45:40.04, DEC = -29:00:28.17). 

$^b$ Values of this column are calculated by subtracting ($Q/I_{{\rm peak}}, U/I_{{\rm peak}}$) = (0.0363, 0.0377) from observed ($Q/I$, $U/I$). 

$^c$ Spectral type and object name from \cite{pau06}.

$^d$ Late/early type from \cite{buc09}.

$^e$ Object name from \cite{vie05}.

$^f$ Based on $KLM$ two-color diagram in \cite{vie05}.

$^g$ Based on $K-L$ versus $K$ CMD in \cite{cle01}.

$^h$ Known bow-shock sources in \cite{tan02, tan05}.

}
\label{tab:fea}

\end{table}
\end{landscape}
\vspace*{6cm}
\begin{table}[h]
\caption{Calculated position angles of bow shock and polarization angles.}
\begin{center}
\begin{tabular}{ccc} \hline
ID           & PA$_{{\rm bow shock}}$$^a$ & PA$_{{\rm polarization}}$$^b$ \\ \hline
2            & 20                         & 170 \\            
3 (IRS\,10W) & 10                         & 120 \\
4            & 30                         & 100 \\
6 (IRS\,21)  & -20                        & 15  \\ 
IRS 1W$^c$   & 5                          & 110 \\ \hline
\end{tabular}
\end{center}
{\footnotesize
$^a$ PA$_{{\rm bow shock}}$ corresponds to the direction of relative velocity vector between the DES and the Northern Arm flow on the plane of the sky. 

$^b$ PA$_{{\rm polarization}}$ corresponds to $\theta_{{\rm int}}$ in Table \ref{tab:fea}. 

$^c$ IRS\,1W is measured in only the first day in our observations.

NOTE.-We cannot obtain position angles of \#9 and \#11 because they exist inside the bar, which is one stream of the minispiral and \cite{pau04} have only the data of the Northern Arm.
}
\label{tab:pa}
\end{table}

\end{document}